\documentclass[journal]{vgtc}                
\ifpdf
  \pdfoutput=1\relax                   
  \usepackage{graphicx}                
  \DeclareGraphicsExtensions{.pdf,.png,.jpg,.jpeg} 
\else
  \usepackage{graphicx}                
  \DeclareGraphicsExtensions{.eps}     
\fi%

\graphicspath{{figures/}{pictures/}{images/}{./}} 

\usepackage{microtype}                 
\PassOptionsToPackage{warn}{textcomp}  
\usepackage{textcomp}                  
\usepackage{mathptmx}                  
\usepackage{times}                     
\usepackage{cite}                      
\usepackage{tabu}                      
\usepackage{booktabs}                  

\onlineid{0}

\vgtccategory{Research}
\vgtcpapertype{please specify}

\title{Visual Companion for Booklovers}

\author{Zona Kostic, Jared Jessup, Jeffrey Baglioni, Nathan Weeks, \\ Johann Philipp Dreessen, Ning Chen, Tianyu Liu \\ \\ Harvard University}

\shortauthortitle{Biv \MakeLowercase{\textit{et al.}}: Global Illumination for Fun and Profit}

\abstract{An innumerable number of individual choices go into discovering a new book. There are unmistakably two groups of book-lovers: those who like to search online, follow other people’s latest readings, or simply react to a system's recommendations; and those who love to wander between library stacks, lose themselves behind bookstore shelves, or simply hide behind piles of (un)organized books. Depending on which group a person may fall into, there are two distinct and corresponding mediums that inform his or her choices: digital, that provides efficient retrieval of information online, and physical, a more tactile pursuit that leads to unexpected discoveries and promotes serendipity. How could we possibly bridge the gap between these seemingly disparate mediums into an integrated system that can amplify the benefits they both offer? In this paper, we present the full-stack redesign of the BookVIS project \cite{bookvis}, focusing on the system's architecture, user interface, user experience, and information design and visualization. BookVIS uses book-related data and generates personalized visualizations to follow users in their quest for a new book. In this new redesigned version, the app brings associative visual connections to support intuitive exploration of easily retrieved digital information and its relationship with the physical book in hand. BookVIS keeps track of the user’s reading preferences and generates a \textit{dataSelfie} as an individual snapshot of a personal taste that grows over time. The app behind this insight is a product of several iterations of design evolution, and includes an efficient image recognition algorithm, system and information architecture model, and mobile data design and visualization. Usability testing has also been conducted and has demonstrated the app’s ability to identify distinguishable patterns in readers’ tastes that could be further used to communicate personal preferences in new “shelf-browsing” iterations. By efficiently supplementing the user's cognitive information needs while still supporting the spontaneity and enjoyment of the book browsing experience, BookVIS bridges the gap between real and online realms, and maximizes the engagement of personalized mobile visual clues.
} 


\CCScatlist{ 
 \CCScat{K.6.1}{Management of Computing and Information Systems}%
{Project and People Management}{Life Cycle};
 \CCScat{K.7.m}{The Computing Profession}{Miscellaneous}{Ethics}
}

\renewcommand{\manuscriptnotetxt}{}

\begin{document}


\firstsection{Introduction}

\maketitle

Bookstores are those mysterious places where people can easily discover knowledge they didn't realize they wanted to find \cite{NYT}. While searching for something of interest, people may accidentally stumble on another book that piques their curiosity, happily immersing themselves in it for a long period of time. However, the situation with the pandemic (COVID-19) enables growing prevalence of web applications, making the concept of discovering the “book beside the book” a less frequent occurrence. Unlike searches performed with web applications, browsing in bookstores is designed to support exploration and unexpected discoveries, and contribute to a larger serendipity. Physically touching a book, pulling it out a stack or off a shelf, and flipping through it is more stimulating than toggling between the tabs of a web browser. 

Due to their irreflexivity and multidirectional nature, e-books have deconstructed some of the functions of traditional books, such as the possibility for interaction and nonlinearity \cite{NYT}. Online bookstores are changing the user’s behaviors \cite{BookWall} and most people still go online only when they have specific targets in mind. However, it has been shown that users prefer to acquire knowledge using all their senses cumulatively \cite{mobile_virtual}, and books lacking other physical attributes such as weight and paper-like texture are less attractive \cite{e_book_AR}. Many booklovers began offering free books in an improvised libraries in front of their homes, inspired by the need to share and browse in more tactile ways. The lock-down of public libraries has greatly enhanced the use of these forms of community involvement. The question is, how can we reduce the interaction with someone's library, and focus on exploring a few books we might really like?

\begin{figure*}
\vspace{-5mm}
\includegraphics[width=\textwidth]{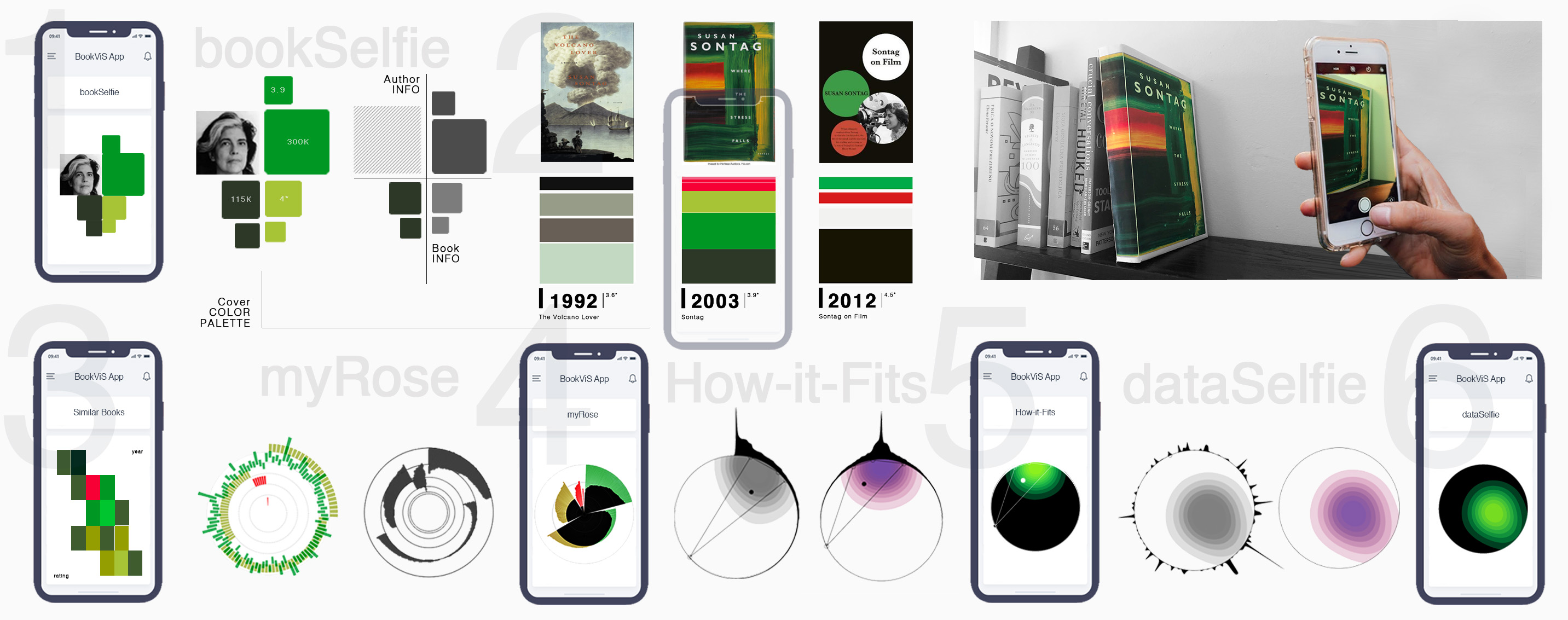}
\vspace{-5mm}
\caption{After taking a photo of a book cover, the \textbf{BookVIS} application generates a set of visual symbols: \textit{bookSelfie}(1), other books from the same author (2), similar books to the book of interest (3), \textit{myRose}(4), \textit{How-it-fits}(5), and \textit{dataSelfie}(6).}
\vspace{-5mm}
\end{figure*}

The majority of book readers believe that e-books will become more popular than printed books in the future, but trending data has not supported this \cite{e_book_printed_book} and many people still prefer reading a physical book. But while this may be true, getting book-related information from the web is increasing because compared to physical data collections where one book is located in a single place, digital libraries have access to infinitely vast amounts of information, and items may be explored from various perspectives. Furthermore, a digital library offers an opportunity to demonstrate multiple relationships between and among books \cite{bohemian}. While some information search tasks \cite{munzner2015visualization} are more easily performed in physical bookstores, digital libraries provide support for more efficient information retrieval. 

In this paper, we are bridging the gap between real and digital realms, utilizing efficient online information retrieval combined with browsing experiences in physical spaces. As browsing for a new book - thought of as an "intellectual stroll" - takes place in environments with distracted or passive attention \cite{PV_PVA}, efficient information retrieval on the fly is crucial. Consumers increasingly supplement their in-store experiences by using mobile phones for fast information retrieval and this is being increasingly utilized in many of the most visited environments at the time of the COVID-19. With the rapid infiltration of portable browsing in shopping, more interactive web experiences will need to evolve, as users are expecting to see more context-based applications. This leads to a challenge of how to maximize the engagement of personalized online experience while keeping readers focused on physical books.

Access to book's data should be easy, consistent, and reliable. Online libraries have become target spaces for the deployment of such applications due to the breadth and complexity of information they can search \cite{mobile_virtual}. However, the current web interfaces are usually too complex and cluttered for users to “browse". With the use of mobile devices being often presented in scenarios with fragmented attention \cite{evaluating_infovis_mobile}, the applications are not designed to provide efficient information delivery. In addition, current visualization interfaces that serve for an easy exploration, discovery, or analysis \cite{munzner2015visualization} only display the results associated with a single attribute, thus requiring users to interact more intensively to find their targets \cite{BookWall}.

Inspired by the experience of a physical bookstore, while utilizing the efficient information retrieval and understanding using smartphones, we developed BookVIS, a web-based application designed with both the physical and digital experiences in mind. The application serves as a visual companion that recognizes book covers and uses them to provide instant visual clues and insight about the books. The application enhances what has become a common habit for most people after spotting a book of interest: taking out their mobile phones and looking for more information online. BookVIS can be easily used on any device which, thanks to good quality rendering, alleviates the need for cell phone research. By snapping a photo and producing salient corresponding book data, we aim to support browsing tasks and enrich users’ experiences. 

BookVIS supports the book-in-hand browsing by generating greater book- and author-related data. This then may be further enhanced by relating it to a user’s stored personal history and preferences for an overview of the user’s unique taste. The book-related visualizations are designed to provide instant information on the spot, leaving space for further exploration if the user is still engaged with the book. Furthermore, visualizations complement “physical" experiences and they are specifically designed for scenarios with distracted attention, such as limited visits. User-related data helps with the creation of a unique image of a personal taste - a \textit{dataSelfie} - that provides further understanding how the book they are holding fits in with their personal preferences. 
 
Moreover, BookVIS also gathers the best practices from User Interface (UI), User eXperience (UX), graphics, and visual communication designs to create an experience that is specifically suitable for portable devices. Instead of a complex visual exploration, we focused on singular book-related data elements, looking for visual solutions that users will be able to grasp in an “on the go" physical space. With the BookVIS project we explore new design methods looking for visual symbolism behind data visualization, transforming a bookstore into a “hybrid space" \cite{hybrid}. We found that this method of unobtrusive visual exploration tools augments users’ perceived sense of place, and enables fast and enjoyable browsing experiences. 

The rest of the work is organized as follows: we present the related work with section 2; System architecture, image recognition model as well as information, UI and UX designs are described in section 3; the subsequent section 4 characterizes surveys conducted with users focusing on qualitative internal and external evaluation, followed by the design evaluation; we discuss the results, limitations, and conclude the paper with future work in section 5.

\section{Related Work}

\begin{figure*}[!h]
\vspace{-5mm}
\includegraphics[width=\textwidth]{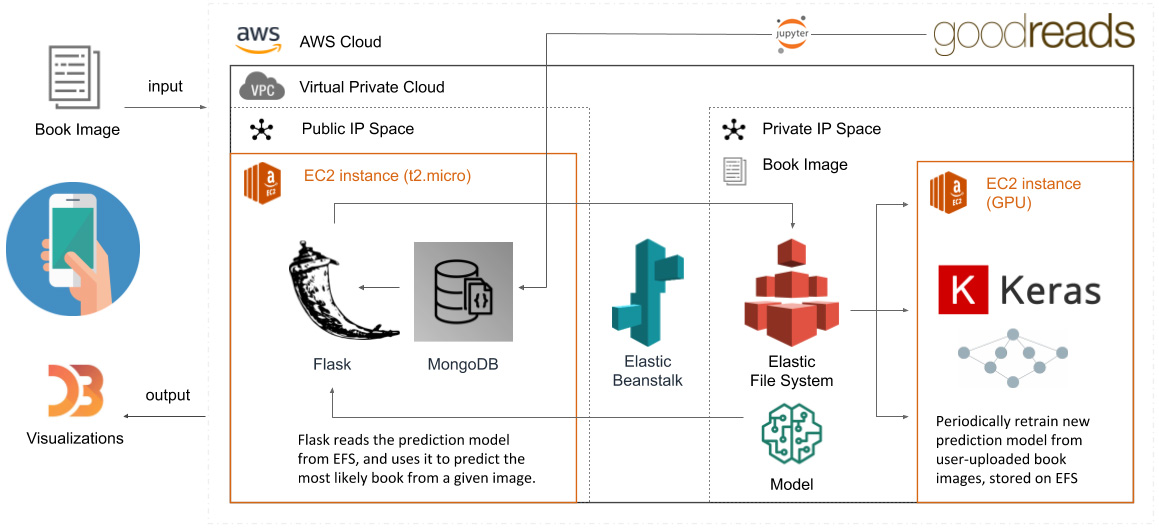}
\vspace{-7mm}
\caption{BookVIS system architecture and its components. The architecture is designed to be simple, dynamic, scalable, and portable.}
\vspace{-5mm}
\end{figure*}

This paper uses the terms \textit{bookstore} and \textit{library} interchangeably, assuming that the object of interest - book - is the same. Unlike in libraries that are more often used for directed information searches, bookstores are spaces where browsing takes place more often. Disregarding the obvious differences, it seems that domain expertise, daily need, or just curiosity would trigger all actions in any of these settings.

Book information search has been mostly performed in online scenarios, supporting efficient information retrieval \cite{bookfish}. However, current interfaces are too complex, narrowing possibilities to a targeted search rather than book exploration \cite{BookWall}. "Shelf-browsing" as a common action in library information search has more traditionally been performed in a physical setting - “library shelves are ideally suited to browsing and they have no electronic equivalent” \cite{blendedShelf}. Online recommender systems support browsing efficiently, but they require prior knowledge of a user’s preferences and/or check-out history. Problems users face when browsing online could be improved when supported by additional more varied and random objects coming from a real-world environment \cite{information_needs}. 

Online systems have been created for more intuitive information search, but they can be too complex for users who still prefer interaction with objects in physical spaces. This inspired some online bookstores to create user-centered interfaces bringing physical bookstore experiences for more efficient target spotting \cite{BookWall}, \cite{best_bookshelf}. However, instead of looking to browse data and understand the entire data corpus \cite{initial_eval}, \cite{analysis_and_vis}, and \cite{sentiment_and_vis}, book lovers browsing in a physical space will generally focus their interest on a single book at the time.

However, getting more on-point book recommendations from a massive selection of books is becoming an interesting and desirable challenge to overcome. Most of the scientific community was focusing on helping users with personalized book recommendation systems, supported by complex visual exploration interfaces \cite{recc_visualization}. This approach has improved information searches in state-of-the art systems for bookstores and libraries \cite{personalized_recc}. Goodreads \cite{goodreads} has been classified as one of the best solutions for personalized information retrieval. Most of the users agree that the Goodreads is the most complete book database, offering a robust application programming interface despite the need for an UI improvement (more about this in the Evaluation section).

Emphasizing the use of digital media in bookstores and libraries was a pioneering approach in bridging the physical and digital arenas. Big panels and screens augment people’s experiences, involve more natural behaviour, and enable on the spot information retrieval. What’s more, the interfaces were designed with users in mind focusing on easy-to-use \cite{info_seeking}, playful \cite{bohemian} \cite{Krogh2004HelpMP}, artistic \cite{info_gallery}, spatially interactive \cite{Allalouf2014VISFACETFV}, and tangible \cite{driving_in} interfaces to support direct contact \cite{AR_in_library}, \cite{mobile_AR} with a book of interest. "Design strategies that combine social, spatial, and digital augment users’ cognitive sense in libraries" \cite{hybrid}. To improve the bookstore or library experience, information about the books should be integrated into the structures that exist in physical settings \cite{putting_book_back}. 

Before the COVID-19 pandemic, libraries were used to target spaces for deployment of digital realities, aiming to provide the users with an immersive experience for consuming resources and related metadata \cite{mobile_virtual}. The immersive systems significantly reduce the time to search for a given target \cite{ar_way}, \cite{hololibrary}. Smart features that are used to augment visualization in book space enhances the reading experience and provides more efficient manipulation with objects in the real-world \cite{enlightwebsite}, \cite{nimble}. Locating, navigating \cite{online_shopping}, \cite{librari}, and keyword searching \cite{book_browse} are supported by virtual reality applications to augment spatial context-driven visualizations. For example, wayfinding has been used in libraries to perform goal-oriented search tasks \cite{Speculative}. Exploring a physical library allows users to choose different paths in a large book collection \cite{blendedShelf} and experience how subsets of books create different visual expressions on a bookshelf \cite{digital_bookshelf}. 

However, as many of these endeavors have relied on specialized virtual-reality technology, not everyone has the opportunity to utilize such devices in everyday life. Unlike digital reality applications, mobile phones have rapidly gained popularity due to their wide range of functionality and portability. The powerful processors and high resolution displays accommodate high-quality graphics with a fine level of detail \cite{comparative_animation}. People routinely consume data via portable devices, from weather forecasting to sleep readings \cite{visualizing_ranges}. Combined with appropriate visual data analysis, they become vital for rapid decision making \cite{network_and_sensor}. In contrast to desktop-based applications for information visualization, mobile devices offer the potential to provide a dynamic and user-centered interface \cite{vis_tiles}. 

Portable devices are available in different screen sizes, offering different types of opportunities for insight, particularly in information visualization as the requirements for anytime/anywhere data access increase \cite{beyond}. Compared to cell phones, tablets provide significant space for augmented reality-based content. However, with tablets users need to use both hands, and our choice in utilizing mobile phones was based on the capability to perform bi-manual interaction with the book and the application simultaneously. Smart watches would be the best option, bridging the gap between hands-free and user-based applications, but they currently lack the device embedded camera. 

The need for personalized information in the form of recommendations is moving beyond current recommendation systems and towards the search for a meaningful representation of the book \cite{book2vec}, while providing visual support in the discovery process. From a bulk of raw data, the idea is to visualize single book-related or personal information in order to construct a taste framework. The user can check her visualization of the personal data and use it as a \textit{dataSelfie} \cite{selfies} to understand and communicate the taste to others.

Designing data visualization for mobile context-based applications that supports physical experiences has many challenges. In the next chapter we will decompose all of them into three groups: web application infrastructure, image recognition modeling, and UX/UI and information design. 

\section{System and Information Architecture}

This section provides a detailed description of components of the BookVIS application. The full-stack redesign of the original BookVIS project \cite{bookvis} is described, focusing on the system's architecture, user interface, user experience, and information design and visualization.

\subsection{System Architecture}

The application architecture (fig. 2) is designed to be simple, dynamic, scalable, and portable. The current iteration has been set up to run on Amazon Web Services (AWS) \cite{aws}, and takes advantage of several technologies offered by that platform, including two EC2 server instances, an Elastic File System, and orchestration via Elastic Beanstalk. 

A dedicated EC2 server instance houses the model developed using Tensorflow and Keras, to support the recognition process. The EC2 instance is powered by a GPU which is able to more quickly compute the model, make it available to the application, and periodically re-train using new images. These may be images that are uploaded directly by users of the application when scanning book covers, but they may also be covered from other data sources to improve the model. This will ensure that the model is capable of recognizing an ever-increasing number of book covers over time. 

The file system is mountable as an Network File System (NFS) volume accessible from all EC2 instances, including both the model and the web application. An Elastic Beanstalk environment is used to orchestrate the web application, and its related resources. Furthermore, a database which includes a more user specific information is required to be persistently maintained. The BookVIS application uses MongoDB database \cite{mongo} for unstructured user-related data storage. 

The user interacts with the application thought the web interface that provides an option to upload an image, and returns resulting information, recommendations, predictions, and most importantly - "d3.js"-based visualizations \cite{d3}. The system is designed to process information on the fly as efficiently as possible, and a portion of the data is retrieved directly from online sources. Once the model "recognizes" a book cover, a point-in-time data call is made to populate information about the book. This information about the user is also accessible via API, with some parts (e.g. users' likes) being stored locally due to inability to access the Goodreads private database.

Additional API calls are made to the user’s Goodreads bookshelves to access custom lists the user has created, that may include books previously read, or any of lists of any possible interest the user may have as a reader. These lists serve as a reference point to assess the compatibility of the current book to the user’s previously saved books. Finally, and utilizing on-the-fly API calls again, users may save books to any of their previously created Goodreads lists, or create a new list, using the application. 

\subsection{Image Recognition Model}

Successful image retrieval should include image preprocessing, image segmentation, extracting image features, or similarity comparison. Previous research in the computer vision field emphasizes research in a content-based image retrieval, classification, and analysis \cite{content-based-retrieval}. In content-based image retrieval and image classification-based models, high-level image features in the form of vectors are extracted and numerical values compared.

In cases when the models’ transformations are known, an effective technique is to extract a scale-invariant or rotation-invariant features, match them, compute the transform, and calculate the distance. We implemented Scalable Vocabulary Tree for Object Recognition (SIFT) as described in this paper \cite{SIFT}. SIFT descriptors extracted from local image regions are hierarchically quantized in a vocabulary tree that allows for more discriminatory vocabulary to be used efficiently as well as integrates indexing and quantization, allocating weights for each SIFT feature used in scoring. When a book cover image is captured, the algorithm will find the cluster that it corresponds to, and then it will be compared with the images in the cluster. After the vocabulary tree was constructed, we took the SIFT descriptors of the image and compared them using the distances with cluster centers values at each node of the tree.

Extracting SIFT features (fig. 3, bottom) proves great for various sorts of cover transformations. With the SIFT technique, the input image was easily extracted from the bookshelf background where the book was placed. However, it is inefficient to retrieve the correct book edition if the same cover design was introduced. Thus we decided to improve it by adding an Optical Character Recognition (OCR) model \cite{ocr}. The OCR served well for phrases like “2nd edition” or retrieving the book data solely based on the image of the book spine.

Despite additional functionalities that OCR brings, some of the limitations are evident. First, geometrical distortions caused by the camera or the photo rotations decrease the chances of good quality text recognition. Second, handwriting typefaces were inaccurately recognized. Finally, most of the “edition” words are on the books’ spines. Unless we start using the image sequence as the input parameter (e.g. the combination of spine and cover images) some information is lost in one-to-one image models like the one used in this work. 

As mentioned, keypoint detectors and feature descriptors are successful in describing local image structures. In computer vision, Convolutional Neural Networks (CNN) models have been proven to be the best performing candidates. Despite the fact that CNNs are generally defined as regressors, making selection problems (like implementing detectors) is very challenging, and recent methods have used CNNs for very fast keypoint detection \cite{bow}. In addition to keypoints and descriptors, the bag-of-words (BoW) CNN models are used for successful implementation of image retrieval applications. 

We used \cite{radenovic}, a CNN\textunderscore{BoW} model as the third model for book cover recognition. After fine-tuning, the model was retrained on the same image training data to learn projections and detectors. The proposed method does not require any manual annotation and takes care of tilted, distorted, or images with poor resolution. Furthermore, this method helped eliminating the need for data augmentation, while still being fast and requiring less memory. 

\begin{figure}[h!]
\centering
\includegraphics[width=230pt]{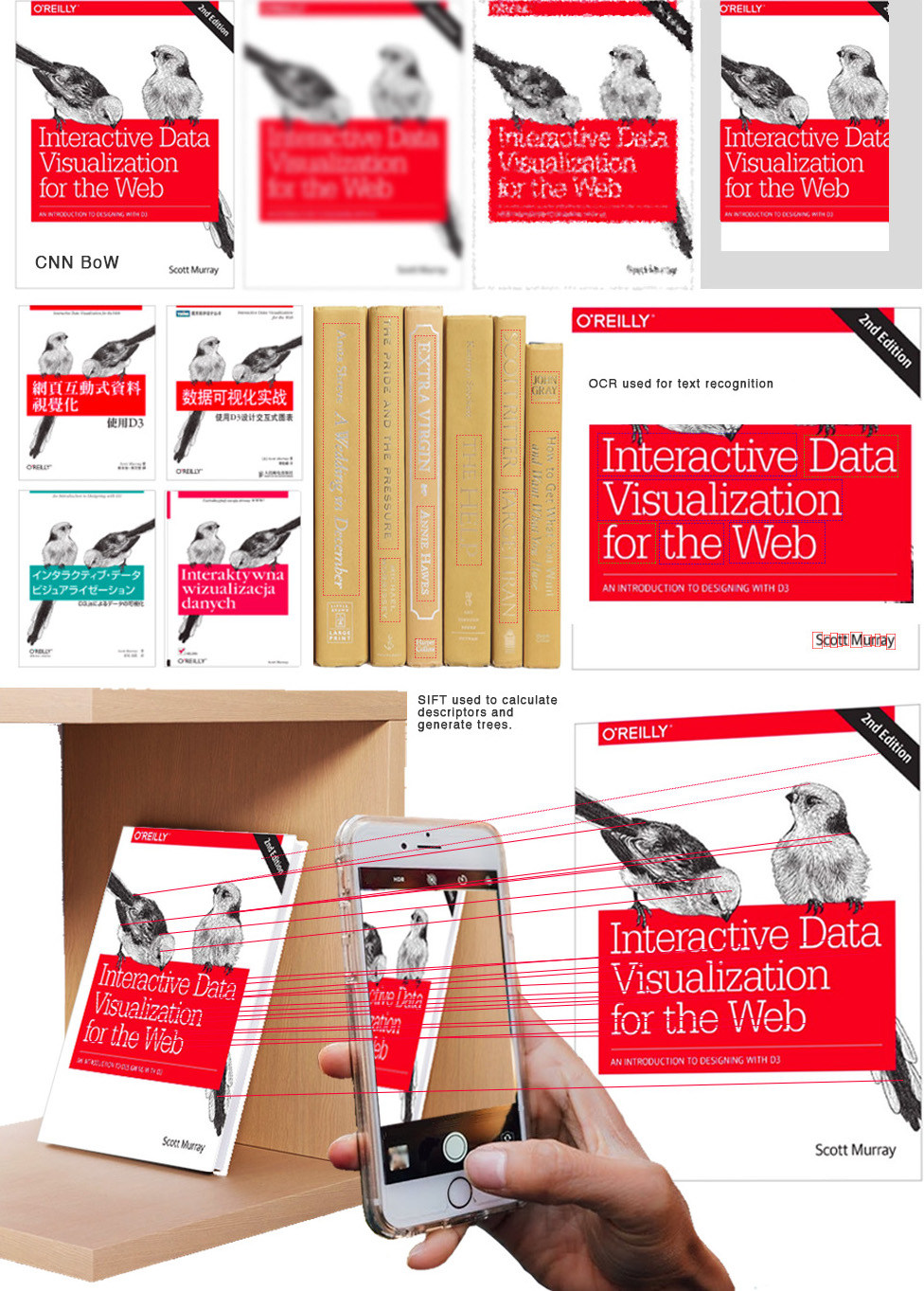}
\caption{Different quality of images of a single book cover (top); Different languages, books' spines, and "2nd edition" detail (middle); The model retrieves correct book information from a tilted query image (bottom).}
\end{figure}

Our final model consists of the combination of the three techniques, SIFT + OCR + CNN\textunderscore{BoW} into an ensemble. Additionally, if a special word like “edition” is detected, it is stored in the database for further processing. Taking images of the books’ spines proved inefficient for unseen images with handwriting typefaces. Our focus was to resolve the most challenging cases like recognizing books from images taken at different angles or retraining the model with poor quality uploaded images from the user. Some additional steps were considered for further model improvement. 

\begin{figure*}
\vspace{-5mm}
\includegraphics[width=\textwidth]{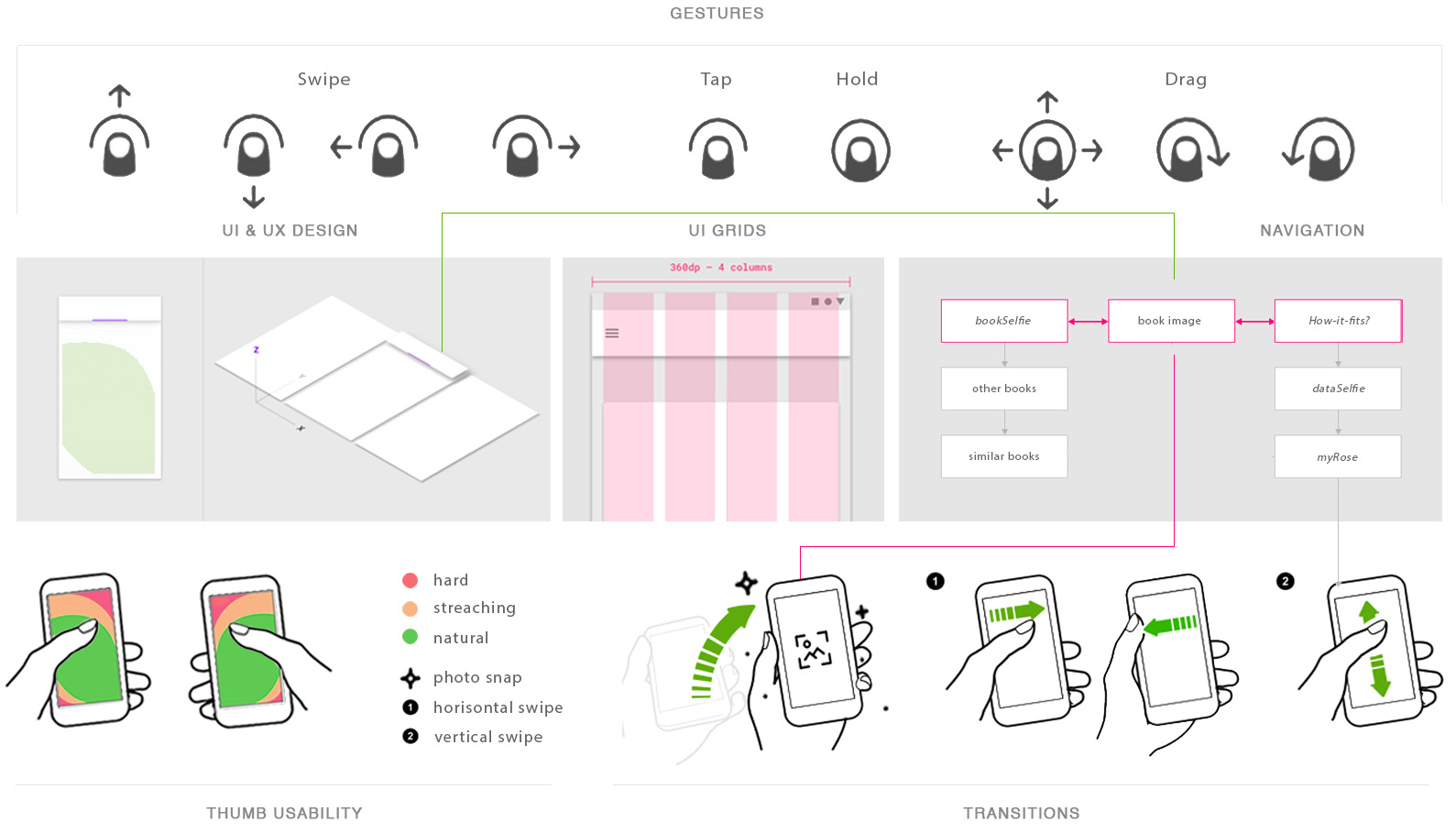}
\vspace{-5mm}
\caption{User Interface and User Experience Design. A set of available gestures has been presented with the top row; UX design to support a "dashboard"-like narrative respecting Material Design guides (middle); thumb usability and transitions are presented with a bottom row illustrations.}
\vspace{-5mm}
\end{figure*}

\subsection{UI/UX and Information Design}

Unlike previous subsections, the design process cannot be translated into linear thinking nor be fully planned in advance. In needs to rely on the constraints imposed by the web application's main concept. The BookVIS design "converts" predominantly textual data into graphics, that should be visually symbolic and meaningful to the user, similar to the way that visual encoding in a lineup of five stars is immediately symbolic of a ranking. We started from an assumption that people can learn a lot about something using associations and quick visual interpretations ("glanceables") will lead to more engagement with the physical objects and data browsing, while repeating the same actions will help understanding its symbolic.

To meet all the mentioned goals, we defined a set of design challenges that helped focus our design process: how to make information easily viewable at a glance; how to provide information that complements the object (a book in hand); how to create personal and visual database with easy to navigate user interface; how to generate a unique image of a personal taste as the novel visual communication language; and how to store a “physical experience” in a digital form. In order to respond to these challenges, the design process needed to be conducted under three separate subdomains: user interface, user experience, and information visualization. 

\subsubsection{User Interface and Experience Design}

Data visualization designers often work across different specialisms, designing the overall user experience, organizing information architecture, and considering the granular representation of the user interface. For an efficient experience, we explored the way to translate users’ needs by considering their journeys and intentions in order to establish an overall aesthetic. All of the components heavily depend on the overall BookVIS concept, composition, device, and data. We are focusing on mobile phones as devices that can be easily manipulated with one hand, while at the same time think of a consistent user interface and enjoyable user experience.

When first introduced, mobile phones enabled an entirely new experience based on innovative applications. Good quality UI and UX design helped accustom users to naturally and intuitively using the devices. Still, some of the disadvantages of mobile phone’s exist, such as thumb size affecting touch points, small screen sizes, no mouse-like precision, no hover functionality, and shorter audience attention spans. Most of the anticipated behaviors from smartphone users is through gestures such as clicking, scrolling, or pinching.  Although there is considerable progress in data visualization, we do not think that we are seeing the full spectrum of possibilities with gestural inputs on mobile applications perfectly integrated with visual representations \cite{not_all_gestures}.

An efficient UI and UX for small screen sizes, limited gestures, and maximum support for the user's physical experience served as a guidance for the design choices, as did browsing over traditional information searches. To address the potential need for use in bookstores, delivering the experience of multiple visual representations was a priority. Some previous research in this space has explored the dashboard-like experience for small screens in the form of the multiple tiles, where each one is like a separate, portable device \cite{vis_tiles}. Using multiple linked charts was demonstrated one way to overcome screen limitation. However, typical interfaces on smartphones usually present thumbnail images or charts in a sorted list. Furthermore. using such a grid-based approach can become tiresome as typically much scrolling is involved to find a specific detail. 

One of the key points of designing for mobile phones is the sense of presence. It is critical for creating a high performing experience. Apart from limited gestures, we needed to consider how transitions are handled from one view to another to provide relevant feedback by the UI components. Taking into account guides summarized with Google Material Design \cite{google} and Adobe Design \cite{adobe}, we decided to take the traditional data dashboard-like organization and transform it into a meaningful narrative of connected pages. Following the predetermined and easy to remember connections between pages, the UI was designed for seamless navigation and an effective UX presence-awareness (fig. 4). While complex desktop applications typically use the concept of nested lists to accommodate this, in our case the user has to be able to access the information instantly and easily on the spot.

One possible approach to create touch-based interactions is to develop touch-traced patterns, which can then be recognized and used to trigger specific interface responses. These touch-gestures can quickly become complex and can be hard to learn and remember. Instead, our goal is to keep the touch interactions simple, with easy manipulation and navigation.

The BookVIS experience design concentrates around both user’s hands at the same time: one will hold the book and another the cell-phone. In such scenarios with limited focus and attention, users have to be able to navigate using only thumb fingers. That narrows down the already limited number of touch gestures we can implement. We decided to use the \textbf{swipe} similar to scrolling as the main gesture to move in between the pages, \textbf{drag} for the hover-like functionality, \textbf{tap} to select, and \textbf{long press} to save. Double-tap has been excluded, due to a very slow manipulation if the phone was used by a non-dominant hand. 

Next, the organization of visualizations has to follow some meaningful order, for the sake of the easy navigation, memorization, and self-awareness. The user experience follows the organization presented with the figure 4, subsection "navigation". The pages will be populated with visualizations, and then, the suggested navigation will be translated into a meaningful visual narrative. The windows are divided into two groups: "left-hand side" pages show the book-related information; "right-hand side" pages are reserved for user-related visualizations. The switch between these two groups can take place with swiping left-right, while navigating in-between the groups takes place with swiping up-and-down.

\subsubsection{Information Design}

Visual exploration is a time-consuming and complex process, and a single analysis session can consist of hundreds of individual steps. Visualizing data on mobile phones is increasingly prevalent in practice, yet it has attracted little attention from the visualization research community \cite{reaching_broader}. Moreover, the users of our applications are not domain experts, making visualizations hard to grasp by the general audience \cite{Q_Selfers}. We lack guidance on how to effectively design visualizations on small displays \cite{visualizing_ranges}. Spanning the contexts of informal data analysis, the BookVIS application specifically considers scenarios in which people encounter visual representations of multivariate data on mobile phones.

\begin{figure}
\centering
\includegraphics[width=\linewidth]{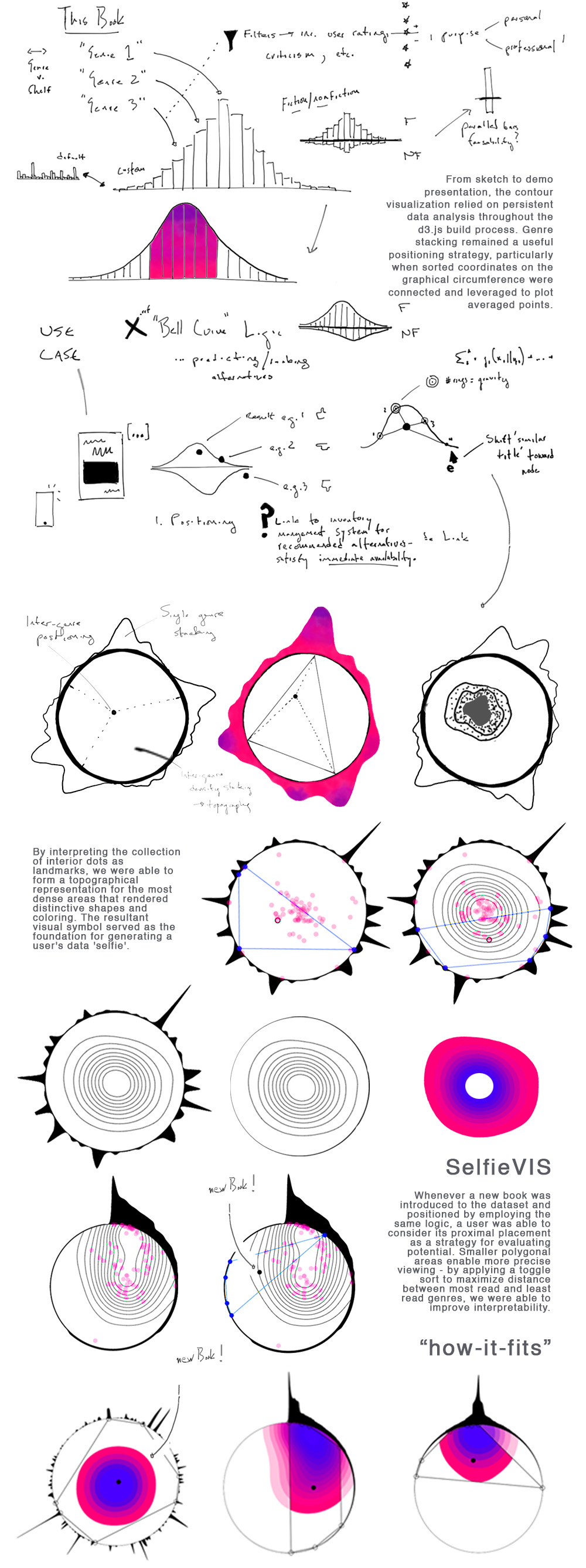}
\caption{Iterations of the \textit{dataSelfie} and \textit{How-it-fits} creation process.}
\end{figure}

Before starting to read a new book, most users would take careful consideration of its genre, ratings, or reviews. Synopsis is also of interest to users, but given that it is usually presented with the book, we excluded it from our variables. Users also like to use applications to save their reading history. After a few consecutive sessions in libraries, if considering digital data (using cell-phones to gain more information) booklovers would most probably like to see:

\begin{itemize}
\setlength\itemsep{-0.5em}
  \item More info about a particular book 
  \item Similar books to a book of interest
  \item Other books from the same author 
  \item Store a book on device for later review
\end{itemize}

Additionally, after being given as an option, users would also like to know if a book is the “right” choice for them, will they like it, or how will it fit their tastes. More about this in the Evaluation section.

Besides having the user interface that helps user’s orientation and promotes effective user experience, the biggest challenge was to reach visualization goals while dealing with the small screen size and limited gestures to present information in a way that doesn't distort it. Furthermore, we wanted to help users navigate data with context that emphasizes comparison. We also wanted to adapt visualizations which anticipate user needs on data depth, complexity, and modality. To overcome these dynamic design challenges, our goal was data visualizations that use custom styled shapes to make understanding on the fly easier, in ways that suit the user’s needs in a given context. Charts can additionally benefit from customizing the graphical elements like composition and typography, and we used our domain experiences to maximize its utilization. Following these new set of constraints, we crafted the three visualization scenarios: (1) learn more about the book in hand (ratings, likes, reviews… etc); (2) Visualize immediate connections such as similar books, popular shelves, or other books from the same author; (3) visualize personal taste in the form of visual data selfies \cite{nam_selfie} based on data from personal shelves, as well as use selfies to visualize how the newly discovered book fits the overall user's taste. Finally, use similar colors to enable easy transitions between the states.

Unlike in some of the previous data selfie research \cite{nam_selfie}, we included additional UI/UX constraints based on the teams’ previous experiences with graphics as well as Google and Adobe design principles. The fact that the vast majority of portable devices interactions last just under 5 seconds \cite{glanceables}, brought an additional design constraint. Unlike with very simple charts, it hasn't been widely determined what people can actually perceive from such “glanceable visualizations” during short periods of time and a limited display space. Thus, the data visualization design had to follow visual symbolism (associativity that symbols like logos bring) rather than complex data exploration for effective information retrieval of custom made charts. At the same time it had to provide visual clarity, bring associativity and personalization, and enable a symbol or a "brand" expression and memorisation. 

As with the UI/UX design, information design choices were determined by the two primary navigation choices (Fig. 4): book-related and user-related sections. Book-related “symbols” are using a book of interest as an anchor point, while user-related “selfies” consider the user's personal shelf’s data. Personal shelves are all books users stored in their libraries, and among them are usually previously read books that users love, course materials or school books, books users would love to read, books the user wrote, or foreign books (including user’s native language or second language of interest). To support the division of book-related and user-related visualizations as well as to back the high-quality UI/UX design to support awareness, data design employed the application of unique visual marks.

The first visualization called a \textit{bookSelfie} as its name implies, should tell something about that particular book, while following the symbolic  pattern (fig. 1). Pattern will always change, but symbolic stays the same (similar how bar-charts change height or length, but the composition of ordered bars stays the same). Taking into account what majority of users would like to see (if interested in that particular book), we decided to visualize the following important components, in a clockwise manner: (1) information about the author; (2) book average ratings and total ratings; (3) book ratings distance (weighted average); and (4) book review’s count. 

Furthermore, in order to maintain the color consistency throughout the application, we picked up and used the same colors as the cover of the book in the user’s hand. This way, the BookVIS application UI will always resemble the book cover, benefiting from a good color choices set by the book cover designers. Some examples of the book colors, covers, and selfies are presented with the image 6. To establish a set of colors that stand out from a perceptual perspective on the book cover, we used the k-means clustering algorithm \cite{k-means}. K-means categorizes a set of data points into ’k’ groups working on effective distance calculations in an unsupervised manner (the k-means python-module communicated directly with Flask application; for more information about the application architecture see Section 3). 

The second visualization shows other books from the same author. According to this research for visualizing temporal data \cite{visualizing_ranges}, bar-like charts are the most effective visualizations for mobile phones. Following this recommendation, we decided to come up with the stacked-barchart timeline pattern. Here, we extracted the cover colors (the top row) and emphasised the use of typography. The visualization follows the carousel principle, putting the book of interest on the timeline and allowing users to scroll left and right to explore more from the same author. Each time the user chooses another book, the background and the overall color scheme changes to the dominant cover colors. If the author has no more than two books (e.g. Tamara Munzner),  “hidden” tiles will compensate for a lack of data (following a good strategy for design choices where data dictates the appearance of visualizations \cite{data_changes}). 

For the “similar books” visualization, we used the matrix organization in which we rescaled the ratings and publication years into 5 x 5 fixed grid. For designing “glanceable visualizations” the intuition about the overall timeline is more important than the accurate scale. The combination of the book ratings and publication years always produces a semi-custom regression visualization. Patterns of this approach always differ for different book collections. Also, the Goodreads API retrieves similar books with almost the same ranges for similar books ratings (between 3.8 and 4.5 stars), assuming users would like to see highly rated books. This enabled us to focus on designing and visualizing other variables. The color scheme is consistent, and the depth of the color approximates the “trust” in rating (how many ratings are there per book). The color itself suggests that books are coming from the same genre. 

On the right side (fig. 4, navigation), there are user-related data and the corresponding visualizations: \textit{How-it-fits}, \textit{dataSelfie}, and \textit{myRose}, following a circle as a guiding visual symbolism. Once the user has taken a photo of the book, the BookVIS application will generate visualizations following the described color scheme. The \textit{How-it-fits} visualization takes saved books from the user’s Goodreads account, stores them in the local app’s database, and generates the \textit{dataSelfie}. Then, the selfie is used as the basis for the \textit{How-it-fits} visualization, to compare the newly discovered book to the rest of the user's library. The final design choice for this visualization is based on the intersections between genres the group belongs to and clipping the inner section of it. Here, we allow for dynamic design to take place, given the inconsistent number of genres associated with a single book. Amid the underlined color, it should always be clear how close or far the newly discovered book is from the user’s taste.

\begin{figure}[h!]
\centering
\includegraphics[width=\linewidth]{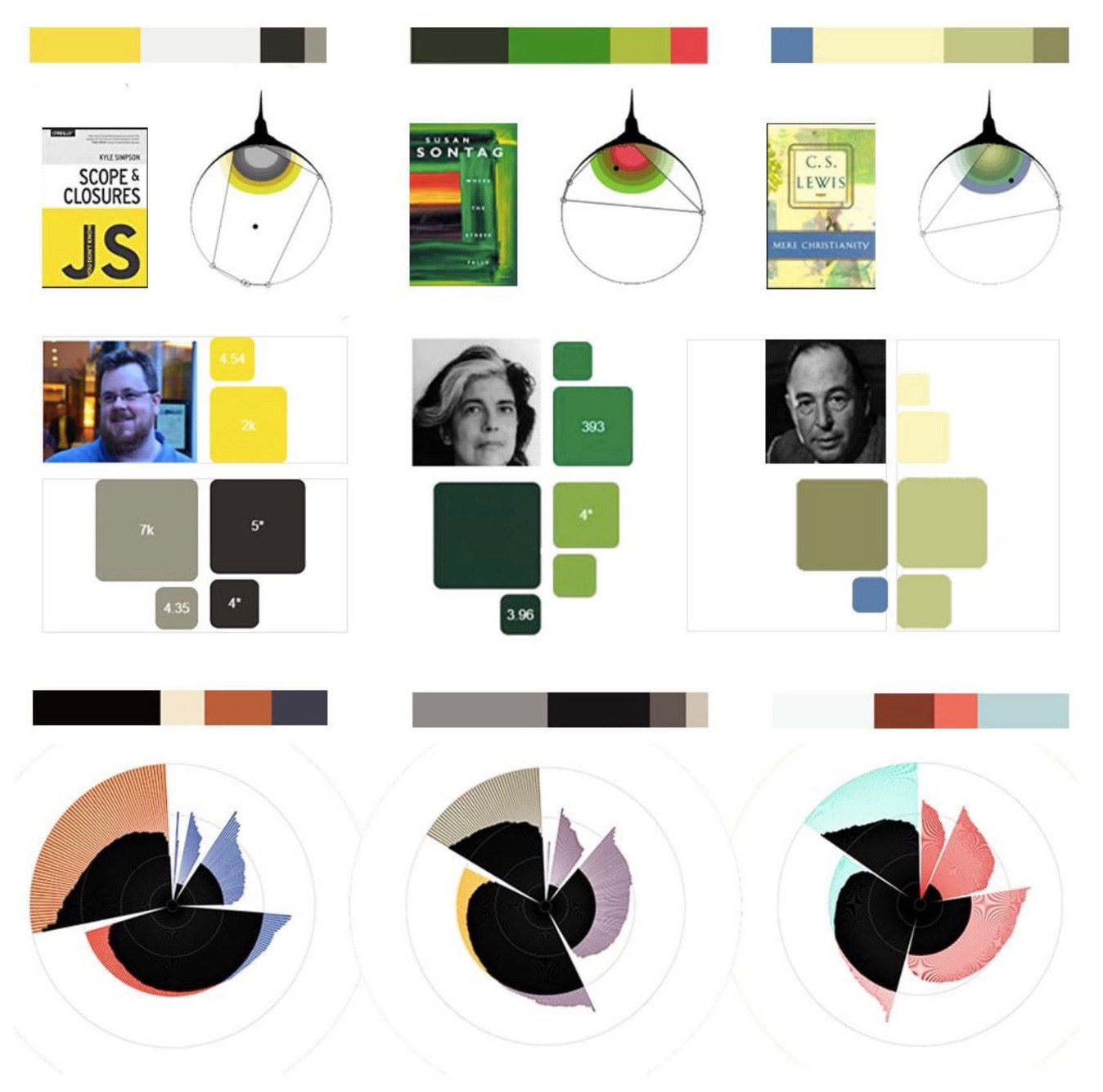}
\caption{From the top to bottom: dominant colors, \textit{How-it-fits} for a single user, \textit{bookSelfies}, dominant colors, and different users \textit{myRose} diagrams.}
\end{figure}

The biggest challenge for us was creating the \textit{dataSelfie}, as the main and leading visualization to illustrate the relatedness of a user’s reading preferences. The idea behind the selfie design started from collecting the genres into histogram (fig. 5). Each book has many genres associated with it thus, piling up and stacking genres would give us the overall idea of the most attractive user’s choices. Book lovers do not necessarily like books coming from one or two genres, more the range of it, and the suggested bell-curved genre distribution was selected to estimate it. This idea was nicely aligned with newly discovered books for \textit{How-it-fits} visualization. A fitness of a new book to the user’s taste can be measured by utilizing visual distances from the genre “peaks” a book belongs to. The book is situated by locating the position of genres it belongs to, constructing the shape that connects them, and placing the dot in the middle of it. Furthermore, the bigger the peak, the “stronger” the force that will move the dot towards it, penalizing the “less” important genres. 

Maintaining a bell-like curve, we employed a circular radial chart to avoid path overlapping and to visually decouple the exterior genre count from the interior multi-genre coordinate position indicators. This idea evolved when dot density was measured and used to form topographic contours with distinct shapes and coloring. The resultant visual symbol serves as a unique personal "logo" or \textit{dataSelfie} for each user and was further optimized for the comparative interpretation of a "book of interest" in the the \textit{How-it-fits} visualization, by toggle-sorting genre count. Sorting became an important first step of undoing the misleading side-effects of alphabetical ordering and for shifting contour alignments away from the circle center. A standard reading would lift high genre books to top of the chart and sink low genre books to the bottom. However, without improved genre hierarchical classification and rating-based genre refactoring, certain test books yielded confusing results for users.

The final visualization, \textit{myRose}, takes the eponymous form of a rose diagram, comparing user ratings and average ratings for each book in a user's library. This strategy highlighted several important trends, including indications of individual taste, 'average' taste, and of the range of a user's critical lean. The Evaluation section shows various patterns encountered from readers with different reading and rating habits. Transitions between visualizations as well as interactions also allow users to immediately perceive that the chart was updated to a revised view and to rebase readings according to anchor variables (e.g., 'user rating' or 'average rating').

\section{Evaluation}

We conducted two different types of surveys: internal and external. First, we used all of our own private book collections to help understand certain design choices, emphasizing obviously different reading tastes between authors. We benefited from unique book collections to help us design selfie symbols. Different personal libraries helped designing visually consistent, yet unique shapes of personal tastes. 

Afterwards, we conducted an external survey with 6 participants. Figure 8 (appendix) shows all the tasks and questions asked, as well as the responses we got from the participants. We worked with these participants to understand the impactfulness of our visualizations. We asked them to evaluate some of the book-related data including the cover color resemblance and the personal data in the form of selfies. 

All participants in the evaluation are experts in the field of information and library sciences and none of them had any experience with data visualization. They provided a large number of book collections stored in Goodreads, which although they highly recommended, also commented on its poor UI. We used their private collections to create personal selfies, rose, and personal taste diagrams, while for the book-related data we picked a single set and passed the same questions to all participants. During the evaluation process, we asked users to emphasise their home libraries in order to measure the importance of suggested visualizations to the quality of a physical experience. Participants were not recruited in a bookstore due to a restricted recruiting timeline.

At the beginning of the process, participants were only informed that we were developing an application for booklovers and bookhunters, but were not aware of its final design, look, nor intent. As such, we felt we got varied responses related to the visualizations’ usability as well as the look of \textit{dataSelfie}, which will be beneficial for the next redesign cycle. For \textit{How-it-fits} visualizations, we used the same 10 books with all participants, combined with their personal book placements for each book to "understand" the taste. Books are coming from different genres, all of them are highly rated, and with similar (and large) number of ratings. We tried to pick the generally “likable” books as these tend to be the books more prominently displayed in bookstores' prime spots, but not necessarily according to everyone's taste.

In general, participants thought it was a nice touch that book-related visualizations followed the colors of the book cover image. Most of the participants (5 out of 6) were able to read and understand the \textit{bookSelfie} visualization. However, stacked-barcharts timeline visualizations were not immediately clear, although the bars with colors were recognized as cover colors. The \textit{dataSelfie} visualization was also not clear at first glance. One of the participants said - \textit{It is showing some trend outside the circle, but I am not sure if there is any relationship between the part that's outside the circle and the dots that are inside}. However, after providing more information about what the \textit{dataSelfie} is - \textit{This is the "look" of all the books coming from your Goodreads library}, participants were able to read it without problems. Most of them expressed the excitement regarding the personal taste visualization idea. 

The effectiveness of the book placement algorithm that generates the \textit{How-it-fits} visualization was highly rated. All 6 participants claimed the new book placement indicator (the "black dot") within a selfie corresponded to their personal taste and well described the probability of "likeness". 


\begin{figure}[h!]
\vspace{-1mm}
\centering
\includegraphics[width=\linewidth]{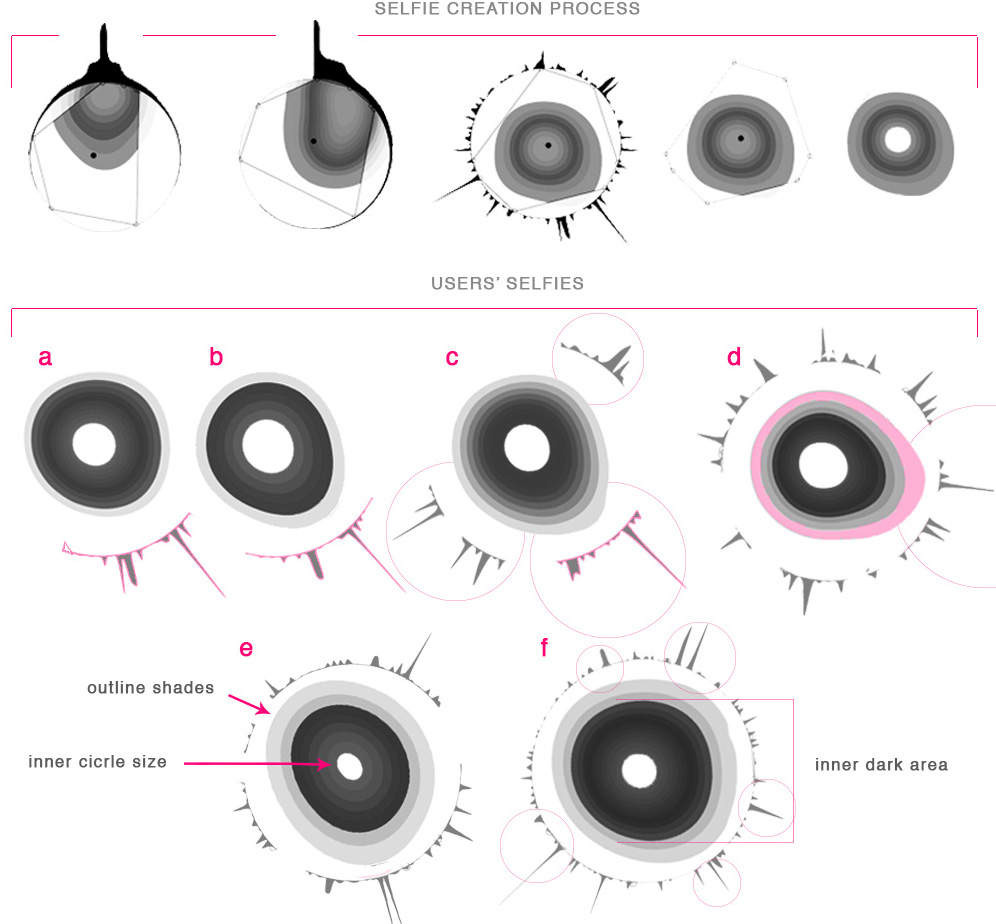}
\vspace{-3mm}
\caption{Subjective analysis and evaluation of \textit{dataSelfies} with designers.}
\vspace{-1mm}
\end{figure}

Lastly, we asked 3 experienced graphics designers to share some thoughts regarding the selfie patterns. We present the summary in the form of a subjective analysis. As we looked more closely into the \textit{dataSelfies} in order to spot obvious similarities or differences, we noticed a couple of “trends” as shown in the fig. 7. Participants “a” and “b” had nearly identical book tastes. That is evidenced by “peaks” concentrated around the same genres (\textit{dataSelfie} is extracted by ordering genres alphabetically). Selfies also show some visual similarity in their construction as well as orientation. For example the number of outline shades as well as the inner black areas are very similar. Furthermore, user “c” also shows a huge interest in books similar to users “a” and “b”. However, user “c” also has a broad interest in other topics. Usually, having interest in many different genres is shown by less “dense” contours, that is present with users “c”, “d”, “e”, and “f”. We could argue that users “a” and “b” have a very clear reading taste, unlike user “d” for example. Users “e” and “f‘ provided very large Goodreads book collections (800 and 1200 books respectively), and the similarity between these two \textit{dataSelfies} is evident in the same inner circle size and two obvious outline shades. 

After observing selfie patterns, we looked more closely at participants' answers (Fig. 8, Appendix), in order to understand the level of excitement and practicality of the application. One of the questions prompt participants to estimate their next most probable move - \textit{You are in a bookstore looking for a new book to buy. After taking a look at a random book and obtaining more information using "glanceable" visualizations, \textbf{what would be your most probable next step}?}. Out of all 6 participants, the user “f” had the most unique response to the question. She stated: \textit{After considering the How-it-fits, the book selected is NOT to according to my taste, so I continue to browse the [physical] shelves.} The user “f” selfie specifically shows the largest inner dark area, indicating a real book-lover with an unspecified preferable genre. After a few "checks" performed with \textit{How-it-fits} visualization, readers gained trust with the placement model that, not only supports their decision process, but also motivates new browsing cycles.

\section{Conclusion and Future Work}

In this paper we have presented the results of the BookVIS application, which keeps track of the user’s reading preferences and generates a \textit{dataSelfie} as an individual snapshot of a personal taste that grows over time. The app behind this insight is a product of several iterations of design evolution, and includes an efficient image recognition algorithm, system and information architecture model, and mobile data design and visualization. Usability testing has also been conducted and has demonstrated the app’s ability to identify distinguishable patterns in readers’ tastes that could be further used to communicate personal preferences in new “shelf-browsing” iterations. 
 
The complexity of such a system highlighted some limitations and many future avenues for improvement. Among them are making comparisons between multiple books, taking pictures of books situated together, and generating multiple overview dashboards. We envision enabling users to simultaneously input more context, i.e., by taking a picture of the entire stack or a bookshelf. More contextual information about the books could be provided, such as the location of the book or similar books within a physical space. These approaches would challenge our smart-phone based visual dashboard, with a new set of contributions provided in the information design, UI, and UX domain. 

After exploring users' surveys, we noticed a couple of edge cases that we didn't take into account and we would love to address them in the application’s future iterations. We focused all visualizations on a single author, but some books have two and more authors. In this BookVIS version, if the book has more than one author, we extracted only the first one. However, the overall visualization concept needs to address this issue. This change will influence not only a redesign of the visualizations, but also components of the system architecture.

As mentioned above, users we tested with seemed to like that the application colors mimic the cover image, but they were unsure how multiple cover designs of the same book would be handled. For the recognition algorithm this is actually a minor issue, but what we are truly interested in is the effect of different covers and colors on differing users’ palettes, perception and cognition, and the subsequent overall “reading” of the BookVIS visual symbols. The dominant color picker doesn't always pick up colors that serve as preattentive visual features. We tested the extraction of more than four dominant colors, but even then, we are not necessarily extracting the most dominant features. Also, more than four colors significantly hardens the coloring of all visualizations. One possible solution to this problem would be using different unsupervised learning techniques or exploring visual saliency techniques and preattentive attention modeling.

Good design promotes unlimited opportunities for further redesign. For BookVIS, the selfies extraction and the insights they produce need more iterations in order to be truly mature. The dynamic redesign - the possibility to encounter for the current as well as the future variables - makes this significantly more complex. Furthermore, we’ll need to run more experiments with a larger and more diverse pool of users to be able to estimate certain selfie shapes and patterns with more confidence. Conducting such experiments demand a well-designed user study with a large number of carefully selected participants.  This is the most probable future path for the next BookVIS redesign. 


\bibliographystyle{abbrv-doi}

\bibliography{main}

\pagebreak
\raggedbottom

\section*{Appendix}

\begin{figure}[!h]
\centering
\includegraphics[width=250pt]{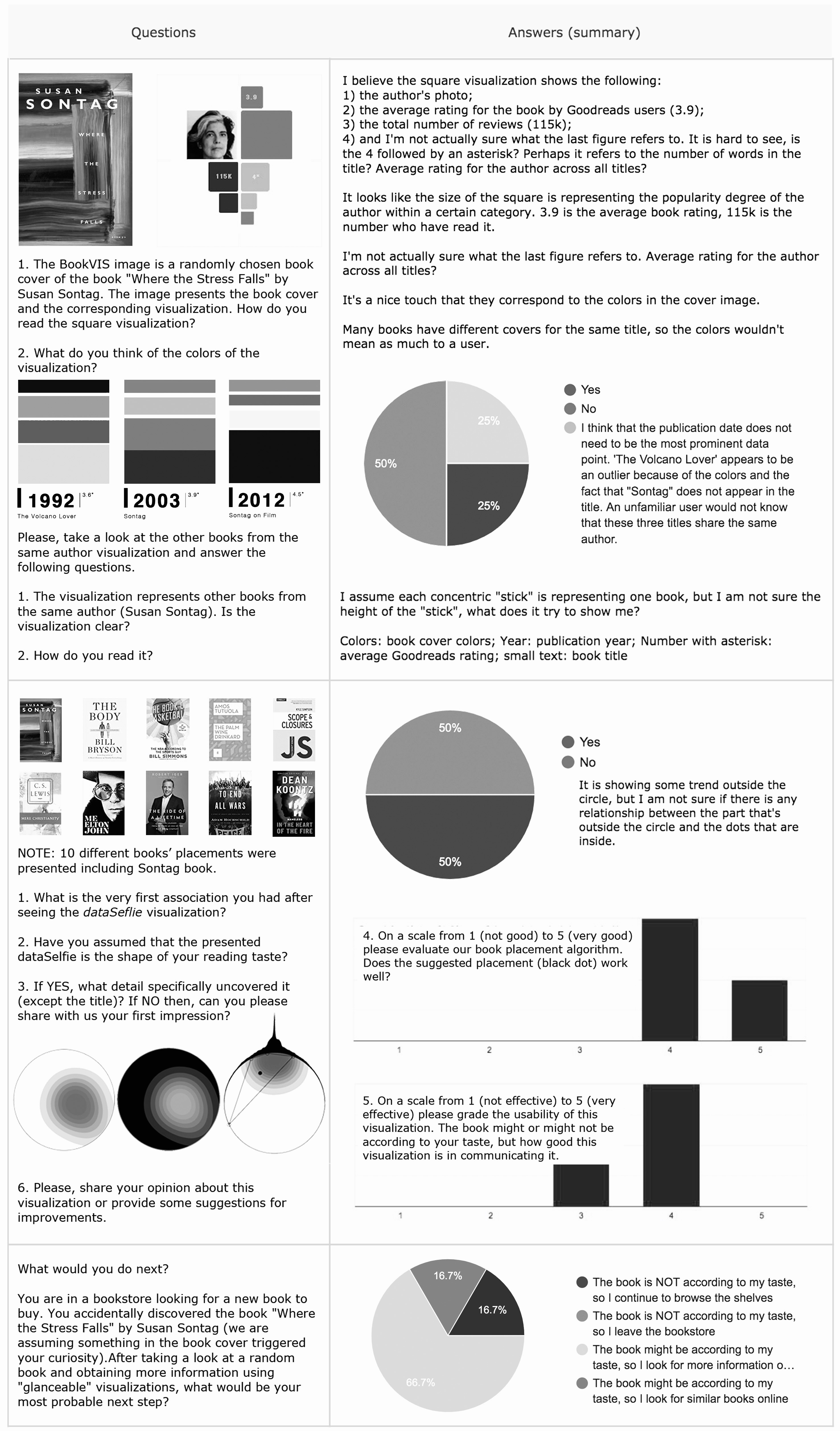}
\caption{Evaluation}
\end{figure}

\end{document}